# INTENSITY ISSUES AND MACHINE PROTECTION OF THE HE-LHC

R. Assmann, CERN, Geneva, Switzerland

*Abstract*

The HE-LHC study investigates the possibilities for upgrading the beam energy of the Large Hadron Collider CERN from 7 TeV to 16.5 TeV. This paper presents a preliminary investigation of intensity issues and machine protection for the HE-LHC.

## INTRODUCTION

The HE-LHC design parameters [1] that are most relevant for collimation and machine protection are summarized in Table 1. It is seen that the total stored energy $E_{stored}$ is 33% higher and the energy density $\rho_e$ is increased 5-fold in each beam. The extrapolation of the HE-LHC is compared in Fig. 1 and Fig. 2 to various accelerators and designs, including the parameters achieved in the LHC during the 2010 run.

The advance in energy density is driven by the decrease in the geometric transverse emittance $\varepsilon_{x,y}$ from 0.5 nm to 0.15 nm. It is noted that the increases of stored energy and energy density are even more pronounced for a single bunch, which must be considered for many machine protection studies.

Table 1: Collimation and protection relevant parameters compared between the nominal LHC and HE-LHC (round beam scenario).

| Parameter | Nominal | HE-LHC |
| --- | --- | --- |
| $E$ | 7 TeV | 16.5 TeV |
| $\gamma$ | 7,461 | 17,587 |
| $\varepsilon_{x,y}$ | 0.5 nm | 0.15 nm |
| $E_{stored}$ (total) | 362 MJ | 482 MJ |
| $\rho_e$ (tot) | 2.9 GJ/mm$^2$ | 15.4 GJ/mm$^2$ |
| $E_{stored}$ (1bunch) | 128 kJ | 242 kJ |
| $\rho_e$ (1bunch) | 1.0 MJ/mm$^2$ | 7.7 MJ/mm$^2$ |

## COLLIMATION EFFICIENCY

The LHC has a sophisticated collimation system [2] that intercepts unavoidable beam losses and safely absorbs them before the associated heat can be deposited in any downstream superconducting magnet. The stored beam energy of $362 - 482$ MJ is to be compared to quench limits of around $5 - 20$ mJ/cm$^3$ in magnets. The collimation system must intercept and absorb stray particles with ultra-high efficiency. The LHC collimation system is located in two dedicated cleaning insertions of the LHC, the betatron collimation system in IR7 and the momentum collimation system in IR3.

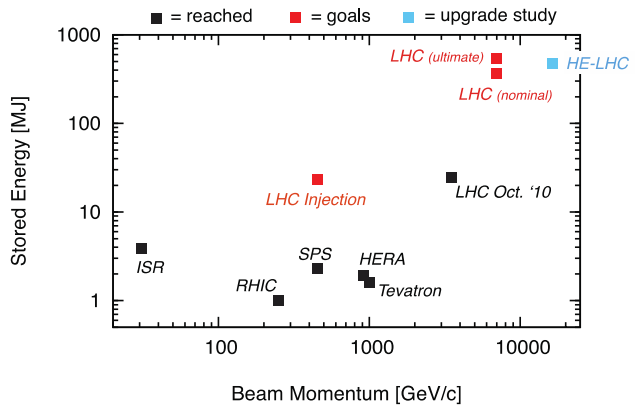

Figure 1: Stored energy per beam versus beam momentum for various accelerators. Filled black squares indicate achieved values, red squares show design values and the blue square represents the HE-LHC design.

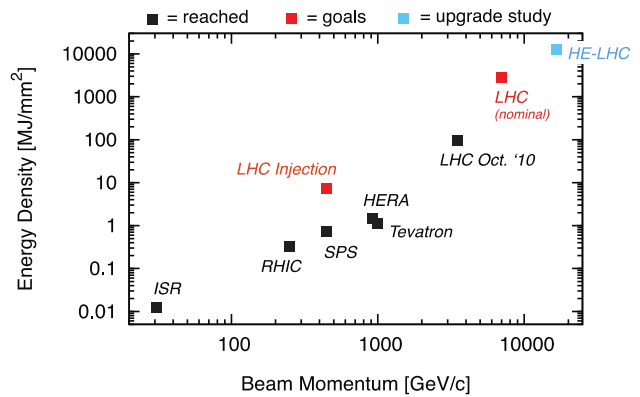

Figure 2: Energy density versus beam momentum. See explanations for Fig. 1.

The LHC collimation system has been designed for optimal performance at 7 TeV along various paths [3]:

- Proper choice of 138 collimator locations for the two beams. 108 collimators have been installed for the first years of LHC operation ("phase 1").
- The use of a 4-stage collimation hierarchy, extending the classical two-stage cleaning design.
- The use of 4 different jaw materials (graphite, fiber-reinforced carbon, copper, tungsten), carefully balancing robustness versus efficiency requirements.
- The use of 2 different lengths of jaws (0.6 m and 1.0 m flat top plus tapering).
- The use of 4 different orientations for optimal coverage in the horizontal (x), vertical (y) and skew planes.

Various nuclear physics processes that depend strongly on beam energy govern the interaction of particles in the collimator jaws. The collimation system therefore behaves differently at higher beam energies.

*Definition of cleaning inefficiency*

The cleaning inefficiency describes the leakage from the collimation system into critical machine elements, for example all superconducting magnets. We define a local cleaning inefficiency as the maximum leakage to one meter of critical superconducting magnets [4]:

$$\tilde{\eta}_{ineff} = \max_i \left( \frac{\Delta N_i / L_i}{N_{impact}} \right) \quad (1)$$

Here, $\Delta N_i$ is the number of lost protons in the superconducting magnet number $i$ of length $L_i$. $N_{impact}$ gives the number of protons that impact on the primary collimators.

*Cleaning inefficiency versus beam energy*

Simulations have shown that the efficiency of the LHC collimation system will be limited by losses in the dispersion-suppressors of the LHC for beams with TeV energies. The energy dependence of the simulated local cleaning inefficiency [5] is shown in Fig. 3 with two possible settings for collimators ("tight" and "intermediate").

It is seen that the LHC cleaning inefficiency gets worse with increased beam energy in the range from 1 TeV to 7 TeV. This is due to reduced multiple Coulomb scattering angles at higher beam energies and an increased probability of single-diffractive scattering.

Single diffractive scattering generates off-energy protons that cannot be intercepted by collimators in the straight sections of the cleaning insertions (lack of dispersive dipole kicks). These off-momentum protons are then lost in the dispersion suppressors downstream of the cleaning insertions. The higher is the beam energy, the higher is the fraction of single-diffractively scattered protons and the higher is the leakage (or inefficiency).

The LHC collimation simulations have been fully confirmed by measured losses downstream of the LHC betatron collimation insertion, as shown in Fig. 4. The proton losses are intercepted, as designed, at the primary collimators. From there onwards, losses are reduced with additional collimators by about four orders of magnitude. Single diffractive protons are lost in two characteristic, superconducting dipoles, as easily seen.

The existing simulation data in the range from 1 TeV to 7 TeV can be fitted as a function of beam energy E, here expressed in units of TeV:

$$\frac{\tilde{\eta}_{ineff}}{10^{-4}} = 0.0276 \frac{1}{m} + 0.0231 \frac{1}{m} E + 0.0051 \frac{1}{m} E^2 \quad (2)$$

This relationship is valid for so-called "tight" collimator settings, referring to nominal settings with primary collimators at 6 σ, secondary collimators at 7 σ, tertiary collimators at 8.4 σ and absorbing collimators at 10 σ.

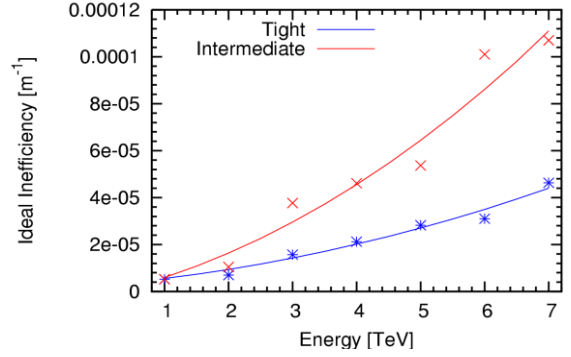

Figure 3: Simulated cleaning inefficiency of the LHC multi-stage collimation system. The two curves show two different settings of collimators. The lines show a fit to the data (see text). The data is from [5].

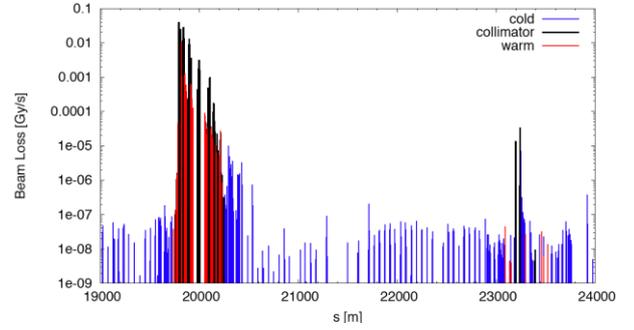

Figure 4: Measurement of proton losses in the betatron cleaning insertion IR7 and through the downstream arc into IR8, performed at 3.5 TeV beam energy. Black bars indicate losses at collimators, red bars at warm machine elements (not critical) and blue bars at superconducting magnets (critical). The beam runs in direction of s.

*Simplified scaling law*

A simplified scaling law can be derived for the probability $P$ of single-diffractive losses versus beam energy. This scaling takes into account the following ingredients:

- The multi-TeV protons traverse an increased integrated length of jaw material. As the multiple Coulomb scattering angle scales with $1/E_1$, more material must be traversed to accumulate enough kick $\theta_{min}$ for reaching the aperture of secondary collimators.
- The required kick $\theta_{min}$ scales with $1/\sqrt{E_1}$.
- The cross section for single-diffractive scattering scales with $\ln(0.3 E_1)$.

Compared to some initial state 0 (with $P_0$ and $E_0$) the impact of single-diffractive scattering scales as follows:

$$\frac{P_1}{P_0} = \frac{E_1 \ln(300 E_1)}{E_0 \ln(300 E_0)} \quad (3)$$

Here, energies are to given in units of TeV. This simplified scaling law is compared in Fig. 5 to the fit from the simulation data. It is seen, that single diffractive scattering can indeed explain the loss of efficiency.

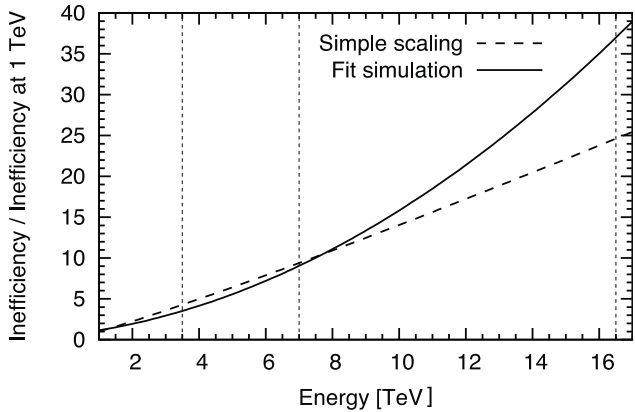

Figure 5: Extrapolation of cleaning inefficiency versus beam energy, comparing a fit of simulation data in the range from 1 TeV to 7 TeV and a simplified scaling law.

According to the two different models it is predicted that the cleaning inefficiency at 16.5 TeV will be increased by a factor between 2.6 and 3.9. This increase in inefficiency (leakage) must be compensated by system improvements in order to avoid collimation-induced intensity limitations. The already foreseen additional collimators in the dispersion suppressors will alleviate this limitation. Detailed studies are required for conclusions of collimation intensity reach at 16.5 TeV beam energy.

## MACHINE ROBUSTNESS

The energy density in the beams and in a single bunch will increase significantly for the 16.5 TeV LHC. The LHC collimators and protection devices have been designed for nominal and ultimate intensities. We assume that all these elements are robust for ultimate bunch intensity and nominal emittance at 7 TeV beam energy. Then we can establish the following brightness limit:

$$\frac{N_p}{\varepsilon} \leq 3.4 \times 10^{20}\ m^{-1} \qquad (4)$$

The present parameters of the HE-LHC study violate this robustness limit by about a factor 2.6. A further study on increased emittance, damage limits or more robust collimator materials is required.

It is interesting that the luminosity reach at the robustness limit is:

$$L \leq \frac{10^{40}(cm\ s)^{-1}}{\gamma \beta^*} \frac{E_{stored}}{500\ MJ} \qquad (5)$$

It is an easy function of the stored energy, of $\beta^*$ and of $\gamma$. The geometric correction factor F from the crossing angle is neglected here.

## ISSUES DUE TO SMALLER GAPS

The primary collimation is set to 5.7 σ in the LHC. Here, we assume that the same normalized setting is required at 16.5 TeV. Due to the adiabatic emittance scaling the absolute half gap of the primary collimators then reduces from about 1.1 mm to 0.6 mm. Transverse resistive wall impedance scales with the third power of the inverse half gap. Consequently, the collimator-induced impedance at 16.5 TeV can be up to a factor 6 larger than at 7 TeV.

An increase of the limiting super-conducting apertures allows relaxing the required normalized setting of the primary collimators. If impedance becomes a limit then it might be required to replace the triplet and other IR magnets with larger aperture hardware.

## HINTS ON MACHINE PROTECTION

The issues for robustness of passive protection collimators have been covered already. There will be additional issues for a few injection and dump protection elements. A detailed analysis by experts is required.

Other possible issues include systematic effects in safety-critical instrumentation, dynamic range limitations in beam loss monitors, interlock thresholds, surveillance levels, etc. A dedicated study by the machine protection experts must address the full picture.

## HINTS ON CLEANING INSERTIONS

The cleaning insertions of the LHC were carefully designed for collimation with the following goals:
1. Establishment of a three stage cleaning per insertion with coverage in horizontal, vertical, skew and momentum phase space.
2. Protection of magnets and accelerator components against excessive heating and radiation damage.
3. Proper radiation control and possibilities for remote handling.

It has to be realized that the available space is already very limited with 7 TeV magnets. The phase advance is at the limit of requirements and cannot be reduced. The optics must be kept similar to the 7 TeV solution. Therefore there is no possibility to decrease the lattice strength, to remove quadrupoles or to increase the beta functions.

The redesign of the cleaning insertions of the LHC for 16.5 TeV is a major challenge.

## CONCLUSION

The parameters of the HE-LHC impose new challenges for operating beams with high intensity:
- A factor between 3 – 6 is lost in collimation efficiency. Improvements must be implemented to compensate this loss. Ongoing collimation upgrades might, however, be sufficient to cope with this.
- The HE-LHC parameters are a factor of about 3 beyond the present robustness limit. Either the emittance is increased or new and more robust materials and technologies should be developed.
- The normalized aperture at 16.5 TeV should be increased by about 50% to avoid operation with small collimator gaps. Such gaps can be operationally unstable and can increase the LHC impedance 6-fold.

Alternatively, new collimator technologies are required.
- Machine protection requires further attention and studies. Presently no show-stoppers are expected.
- The re-design of the LHC cleaning insertions for 16.5 TeV is a major challenge and must be addressed early on in the design process.

## REFERENCES

[1] F. Zimmermann. These proceedings.

[2] R. Assmann et al, "The final collimation system for the LHC". EPAC06. LHC-Project-Report-919.

[3] R. Assmann, "Collimation for the LHC High intensity beams". Proc. 46th ICFA Advanced Beam Dynamics Workshop on High-Intensity and High-Brightness Hadron Beams (HB2010). Sep. 27 – Oct 1 2010, Morschach, Switzerland.

[4] R. Assmann, "Collimators and Cleaning, Could this Limit the LHC Performance?". CERN-AB-2003-008 ADM (2003).

[5] C. Bracco, "Commissioning Scenarios and Tests for the LHC Collimation System". CERN-THESIS-2009-031 EuCARD-DIS-2009-004. Lausanne: École Polytechnique Fédérale de Lausanne, 2009.